\documentclass[12pt]{article}
\usepackage{amstex,epsf}
\renewcommand{\baselinestretch}{1.24}
 
\newfont{\bit}{cmbxti10 scaled 1728}

\def\th{\tilde{h} }
\def\d{\partial }
\begin{document}
\renewcommand{\thefootnote}{\fnsymbol{footnote}}
\newpage
\pagestyle{empty}
\begin{center}
{\LARGE {Symmetries of pp-Waves \\
 with\\
Distributional Profile\\
}}

\vspace{1cm}
{\large
 Peter C. AICHELBURG
 \footnote[1]{ e-mail: pcaich @@ pap.univie.ac.at}
  \footnote[4]{supported in part by the FUNDACION FEDERICO}
}\\
{\em
 Institut f\"ur Theoretische Physik, Universit\"at Wien\\
 Boltzmanngasse 5, A - 1090 Wien, AUSTRIA
 }\\[.5cm]
{\em and}\\[.5cm]
{\large Herbert BALASIN
\footnote[2]{e-mail: hbalasin @@ email.tuwien.ac.at}
\footnote[3]{supported by the APART-program of the
Austrian Academy of Sciences}
}\\
{\em
 Institut f\"ur Theoretische Physik, Technische Universit\"at Wien\\
 Wiedner Hauptstra{\ss}e 8--10, A - 1040 Wien, AUSTRIA
}\\[.5cm]
\end{center}
\vspace{0.5cm}

\begin{abstract}
We generalize the classification of (non-vacuum) pp-waves \cite{JEK}
based on the Killing-algebra of the space-time by admitting
distribution-valued profile functions. Our approach is
based on the analysis
of the (infinite-dimensional) group of ``normal-form-preserving''
diffeomorphisms.

\end{abstract}

\rightline{UWThPh -- 1995 -- 30}
\rightline{TUW 95 -- 20}
\rightline{October 1995}

\renewcommand{\thefootnote}{\arabic{footnote}}
\setcounter{footnote}{0}
\newpage
\pagebreak
\pagenumbering{arabic}
\pagestyle{plain}
\renewcommand{\baselinestretch}{1}
\small\normalsize
\section*{\Large\bit 1) Introduction}
In 1960, in a now classical paper, Jordan, Ehlers and Kundt (JEK) \cite{JEK}
gave a complete classification for the special class of gravitational waves
with parallel rays ( pp-waves) in terms of their symmetries.
These vacuum space-times which admit a non-shearing, non-twisting, and
non-expanding null congruence can be characterized by the existence of a
covariantly constant (cc) null vector field.
In a recent paper the present authors \cite{AiBa} remarked that the JEK
classification fails for the ``gravitational field of a mass-less particle''.
This so-called AS-metric can be obtained by considering the ultra-relativistic
limit of the Schwarzschild geometry \cite{AS,BaNa3}. In the limit one is
left with a pp-wave whose profile-function is concentrated on a null
 hyper-plane. (Whereas in general a delta-like behavior of some of the metric
components leads to ill-defined quantities such as the Riemann-tensor, this
is not so for the case under consideration.) A comparison of the AS-metric
with the JEK-classification shows that this pp-wave should have only two
Killing symmetries. By direct calculation however, it is easy to show that
there are in fact four Killing vectors, as one would expect on physical
grounds. For a detailed discussion see \cite{AiBa}. This discrepancy can be
explained by the fact that in the JEK-classification  the wave profile
was tacitly assumed to be a classical (regular) function rather than a
distribution.

Motivated by this we reconsider the JEK-classification by allowing
for distributional wave profiles.
Moreover, we allow for non-vacuum space-times (see also \cite{SiGo}),
i.e. we do not impose any field equations from the outset.
As a result we show that this generalization does indeed lead to additional
symmetry groups for pp-waves, even in the vacuum case of which the
AS-metric is one example.

\section*{\large\bit 2) Normal-form-preserving diffeomorphisms
and the characterization of pp-waves }

Since we do not want to impose Einstein's vacuum equations we define
pp-waves by requiring the existence of a vector-field $p^a$ and
a 2-vector-field $F^{ab}$ such that $p^{[a}F^{bc]}=0$. A metric will
be called pp-wave if $p^a$ is null and $p^a$ and $F^{ab}$ are mutually
orthogonal and covariantly constant.
In \cite{JEK} it was shown, using adapted coordinates $(u,v,x^i)$,
that any such pp-wave can be written in normal form
\begin{equation}\label{ppWav}
ds^2 = -du\, dv + \delta_{ij}dx^i dx^j + f(u,x) du^2,
\end{equation}
where the cc vector is given by $\d_v$ and the $x^i, i = 1, 2$ are coordinates
on the flat $u = const$, $v = const$ surfaces $\cal{S}$.
However, for a given pp-wave the normal form does not fix the
coordinate system uniquely. Transformations that do not change the form of
the metric (\ref{ppWav}) will be called normal-form-preserving (nfp).
 This subgroup of the full diffeomorphism-group consists of
transformations of the form
\begin{align}\label{NFP}
&\tilde{u} = au + b\qquad\tilde{v} =
        \frac{1}{a}\left (v + 2(\delta_{ij}\Omega^i{}_k
        x^k {d'}^j(u) +n(u))\right )\nonumber\\
&\tilde{x^i} = \Omega^i{}_j x^j + d^i(u)
\end{align}
under which the profile function is changed into
\begin{multline*}
\tilde{f}(u,x) = a^2 f(au+b,\Omega\, x+d(u)) + \delta_{ij} {d'}^i(u)
{d'}^j(u)\\
        - 2 ( n'(u) + \delta_{ij} \Omega^i{}_k x^k {d''}^j(u)),
\end{multline*}
where $a$ and $b$ are constants, $\Omega$ is a constant rotation matrix, i.e.
$\Omega^t\Omega = id$, $d^i(u)$ denotes a vector tangential to $\cal{S}$
depending on $u$ and $n(u)$ is a scalar function of $u$. Prime
denotes differentiation with respect to $u$.
Thus, nfp-diffeomorphisms relating equivalent pp-waves form an infinite
dimensional  group with ``parameters''
$(a,b,\Omega^i{}_j;d^i(u), n(u))$.
The main difficulty in classifying  pp-waves in terms of symmetries is
finding a ``canonical'' representative for the profile functions. Since we
would like to consider distributional wave profiles, it is natural to
restrict to $C^\infty$-diffeo\-mor\-phisms.
(For a general definition of distributions on an arbitrary manifold,
see \cite{BaNa1}.) Note that the nfp-transformations
comprise possible isometries as a finite-dimensional subgroup, namely
as those transformations that do not change the profile function.
Let $(\alpha,\beta, \omega^i{}_j; \theta^i(u),\nu(u))$ be the corresponding
parameters of an infinitesimal nfp-transformation. As a consequence the
Killing-vectors $\xi$ may be written as
\begin{align}\label{Killing}
&\xi^u = \alpha u +\beta,\nonumber \\
&\xi^v = -\alpha v + 2(\theta'(u)\cdot x + \nu(u)),\nonumber \\
&\xi^i  = \omega^i{}_j x^j + \theta^i(u),
\end{align}
and the Killing condition restricts the profile to
\begin{equation}\label{GenKill}
2\alpha f + (\alpha u + \beta)f'  - x\omega\d f + \theta(u)\d f -
        2(\nu'(u) + \theta''(u)\cdot x)=0.
\end{equation}
{}From now on we will suppress the two-dimensional indices,
using a matrix-type notation instead.

\section*{\large\bit 3) Classification of Adjoint Orbits }

Since the nfp-group does not only act on profile functions $f$
but also on the space of (possible) Killing-vectors $\xi$, the latter
action may rephrased in a purely group-theoretic setting without
any reference to a specific profile.
With regard to the notation of
the previous chapter an arbitrary element $g$
of the nfp-group $G$ will be parametrised by
$$
g=(a,b,\Omega; d(u), n(u)).
$$
Composition of nfp-transformations allows us to define the multiplication map
$\mu: G\times G \to G$ in parameter-space together with the adjoint action
$k_g : G\to G$
\begin{align}
g_3 &= \mu(g_2,g_1)=g_2\cdot g_1 \nonumber\\
&a_3 =a_2 a_1,\qquad b_3 = a_2 b_1 + b_2, \qquad \Omega_3 =
         \Omega_2 \Omega_1, \nonumber\\
&d_3(u) = \Omega_2 d_1(u) + \phi_1^* d_2(u),\nonumber\\
&n_3(u) = n_1(u) + a_1 \phi_1^* n_2(u) +
        a_1 (\Omega_2d_1(u))\cdot (\phi_1^*{d'}_2(u))\nonumber\\
&\text{}\\
&\text{where }\phi(u) := a u +b \quad (\phi^*f)(u) := f(au+b).\nonumber
\end{align}
\begin{align}
\th &= k_g(h) := gh g^{-1}\nonumber\\
&a_{\th} = a_h,\qquad b_{\th} = a b_h - a_h b + b, \qquad
         \Omega_{\th} = \Omega_h,\nonumber\\
&d_{\th} (u) = \phi^{-1*}\left( -\Omega_h d(u) +
        \Omega d_h(u) + \phi^*_h d(u)\right ),\nonumber\\
&n_{\th} (u) = \frac{1}{a}\phi^{-1*} \left[ n_h(u) -
        (n(u) -d(u)\cdot d'(u)) + a_h \phi_h^* n(u)+ \right.\nonumber\\
&\hspace*{1.5cm}\left. a_h (\Omega d_h(u))\cdot
        (\phi_h^* d'(u)) -(\Omega_h d(u))\cdot
        (\Omega {d'}_h(u) + a_h \phi_h^* d'(u)) \right ]. \nonumber\\
\end{align}
In order to find the adjoint representation of $G$ on its Lie-algebra,
we consider an arbitrary curve through the identity
$e=(1,0,id; 0,0)$ of the group and calculate its derivative at $e$.
Parametrising a generic Lie-algebra-element by
$$
X=(\alpha,\beta,\omega;\theta(u),\nu(u))
$$
the adjoint representation becomes
\begin{align}\label{GenAdj}
\tilde{X}&=Ad(g) X,\nonumber\\
&\tilde{\alpha}=\alpha,\qquad\tilde{\beta}=a\beta -b\alpha,
        \qquad\tilde{\omega}=\omega,\nonumber\\
&\tilde{\theta}(u) = \phi^{-1*}\left(\Omega \theta(u) -\omega d(u) +
        (\alpha u + \beta) d'(u)\right ),\nonumber\\
&\tilde{\nu}(u) = \frac{1}{a}\phi^{-1*}\left[\nu(u) + \alpha n(u) +
        (\alpha u + \beta) n'(u)+ (\Omega \theta(u))\cdot d'(u)-
      \right.\nonumber\\
&\hspace*{1.4cm}\left. d(u)\cdot(\Omega \theta'(u)+
        \alpha d'(u)+(\alpha u + \beta )
        d''(u)) - (\omega d(u))\cdot d'(u)\right].\nonumber\\
\end{align}
The explicit form (\ref{GenAdj}) of the adjoint representation allows
 us to classify its orbits.
It turns out that there are eight different orbits. From each orbit we may
obtain a particularly simple (canonical) representative
by taking advantage of the adjoint transformations.
 We then require this representative
to be the generator of an isometry. By solving the corresponding Killing
equation the form of the profile function is obtained.
The main advantage of this method lies in splitting the problem into a
profile-dependent and independent part.
 We summarize our results in the following table.
\newline

\noindent
\begin{tabular}{ll}
a) $\omega\neq 0,\alpha\neq0$
        &$X=(1,0,\omega_1;0,0)\quad
        \omega_1=\frac{\omega}{\alpha}\quad$
        \\
        &$uf' +2 f = x\omega_1\d f \Rightarrow
        f(u,x)= \frac{1}{u^2}\Phi(e^{-\omega_1 \log u} x)$\\
&\\
b) $\omega\neq 0,\alpha=0,$
        &$X=(0,1,\omega_1;0,0)\quad
        \omega_1=\frac{\omega}{\beta}\quad $
        \\
\hspace*{0.5cm}$\beta\neq 0$
        &$f'= x\omega_1\d f \Rightarrow
        f(u,x)= \Phi(e^{-\omega_1 u} x)$\\
&\\
\end{tabular}

\noindent
\begin{tabular}{ll}
c) $\omega\neq 0,\alpha=0,$
        &$X=(0,0,\omega_1;0,\nu(u))\quad
        \omega_1=\omega\quad$
        \\
\hspace*{0.5cm}$\beta=0$
        &$2\nu'+ x\omega_1\d f = 0\Rightarrow
        f(u,x)= F(u,x^2)\qquad \nu=\nu_0$\\
&\\
d) $\omega=0,\alpha\neq 0,$
        &$X=(1,0,0;\theta_1,0)\quad
        \theta_1=\frac{\theta(0)}{\alpha}\quad$
        \\
\hspace*{0.5cm}$\theta(0)\neq 0\,\,$
        &$uf' + 2f= -\theta_1 \d f \Rightarrow
        f(u,x)=\frac{1}{u^2} \Phi(x-\log u \theta_1)$\\
&\\
e) $\omega=0,\alpha\neq 0,$
        &$X=(1,0,0;0,0)\quad$
        \\
\hspace*{0.5cm}$\theta(0)=0$
        &$uf' + 2f= 0 \Rightarrow
        f(u,x)= \frac{1}{u^2}\Phi(x)$\\
&\\
f) $\omega=\alpha=0,\beta\neq 0$
        &$X=(0,1,0;0,0)\quad$
        \\
        &$f' = 0 \Rightarrow
        f(u,x)= \Phi(x)$\\
&\\
g) $\omega=\alpha=0,$
        &$X=(0,0,0;\theta(u),\nu (u))\quad$
        \\
\hspace*{0.5cm}$\beta=0,\theta(u)\neq 0$
        &$-\theta(u)\d f + 2\theta''(u)\cdot x
        + 2\nu'(u)=0$\\
&\\
h) $\omega=\alpha=0 $&
        $X=(0,0,0;0,\nu(u))\quad $
        \\
\hspace*{0.5cm}$\beta=\theta(u)=0$&$\nu'= 0 \Rightarrow
        f(u,x)= f(u,x)$
\end{tabular}
\newline

\noindent
Different orbits are characterized by the invariant group parameters
$\alpha$ and $\omega$ and split into sub-cases depending on the other
parameters.
The representative $X$ of each orbit, the Killing equation and the
corresponding form of the profile function  $f$ are given. The explicit
form of the Killing fields is obtained by inserting $X$ into (\ref{Killing}).
Case g) may be classified further with respect to the number
and the degree of the zeros of $\theta(u)$. The simplest case, $\theta(u)$
without any zeros, reproduces the result of \cite{AiBa} (after imposing vacuum
equations),
whereas the others allow concentrated (delta-like) contributions to the
profile. Finally, case h) is the generic  pp-wave  with  one Killing vector,
without any restriction on the profile $f$. Thus each class
(a--g) admits $\d_v$ as Killing vector.
 Symmetry groups of three or more
parameters can be obtained by combining the above cases. Note however,
that this is not straightforward since it is not guaranteed that the Killing
vectors can be reduced to their canonical form simultaneously. A complete
analysis of this problem will be published in an more elaborated paper. Here
we restrict ourselves to discuss pp-waves with delta-like profiles.

\section*{\large\bit 4) Impulsive pp-waves}

Let us focus our attention on profiles of the form
\begin{equation}\label{Shock}
f(u,x)=\delta(u) f(x),
\end{equation}
since they are natural candidates for a richer symmetry structure.
In order to classify these space-times we follow exactly the same path
as in the general case:  We consider those nfp-transformations,
which preserve (\ref{Shock}). Following (\ref{GenKill}) the action of an
arbitrary nfp-transformation on (\ref{Shock}) yields
\begin{multline}
\tilde{f}(x,u)=a^2 \delta(au+b) f(\Omega x +d(u))\\
        +d'(u)^2 - 2(n'(u) + \Omega x \cdot d''(u))=:\delta(u)\tilde{f}(x).
\end{multline}
The $C^\infty$ nature \footnote{Actually, the result remains unchanged if
one admits ``distributional'' ($C^0$) coordinate-changes} of
the nfp-transformations and the location of the
singularity require $d''(u)=0,\> 2n'(u)- d'(u)^2=0,\> b=0$. Therefore
the nfp-group is cut down to a finite-dimensional subgroup $G_0$
\begin{align}\label{rnfp}
&g=(a,0,\Omega; d_0+ u d_1, 1/2 u d_1^2 + n_0),\nonumber\\
&X=(\alpha,0,\omega; \theta_0 + u\theta_1 , \nu_0),\nonumber\\
&\tilde{f}(x) = a f(\Omega x + d_0),
\end{align}
which will be called restricted normal-form-preserving (rnfp) group.
An immediate consequence of (\ref{rnfp}) is the fact that $\tilde{f}$
does not depend on $d_1$. This implies that all impulsive waves admit a
three-parameter Killing-group in contrast to the one-parameter group for
general profiles.
The adjoint representation of the rnfp-group is easily derived from
(\ref{GenAdj})
\begin{align}\label{rAdj}
&X=(\alpha,0,\omega;\theta_0 + u\theta_1, \nu_0) \qquad \tilde{X}=Ad(g)X
\quad g\in G_0\nonumber\\
&\tilde{\alpha}=\alpha,\quad \tilde{\omega}=\omega\nonumber\\
&\tilde{\theta_0}=\Omega\theta_0 - \omega d_0\quad
        \tilde{\theta_1}=\frac{1}{a}(\Omega \theta_1 -\omega d_1 + \alpha d_1)
        \nonumber\\
&\tilde{\nu_0}= \frac{1}{a}(\nu_0 + \alpha n_0 +\Omega\theta_0\cdot d_1
        - d_0\cdot \Omega \theta_1 -\omega d_0\cdot d_1 -\alpha d_0\cdot d_1)
\end{align}
Taking into account that $\theta_1$ and $\nu_0$ are already Killing
parameters, i.e. that they do not change the profile, it is possible
to simplify (\ref{rAdj}):
\begin{align}
&X=(\alpha,0,\omega;\theta_0 , 0) \qquad g = (a,0,\Omega;d_0,0) \nonumber\\
&\tilde{\alpha}=\alpha,\quad \tilde{\omega}=\omega,
        \quad\tilde{\theta_0}=\Omega\theta_0 - \omega d_0\quad
\end{align}
And finally the Killing-equation is reduced to
$$
0=\alpha f(x) - x\omega \d f(x) + \theta_0 \d f(x),
$$
which  restricts the profile $f(x)$.
The analogous classification to (a)--h)) is now
obtained in a straightforward way:
\begin{itemize}
\item[1)] $\omega\neq 0, \alpha=0$: Using the adjoint action (\ref{rAdj})
        it is possible to reduce the representative to
         $X=(0,0,\omega;0,0)$, turning the Killing-condition into
        $x\omega \d f(x) =0 \Rightarrow f(x)=g(x^2).$
\item[2)] $\omega= 0, \alpha\neq 0$: Since the Killing-vector is unique
        up to scalar multiple, we set $\alpha=1$, which leaves us with
        $X=(1,0,0;\theta_0,0)$ and
        $(\theta_0\d)f = -f \Rightarrow f(x) = h(\tilde{\theta}_0x)
        e^{-\frac{\theta_0x}{\theta_0^2}}$,
        where $\tilde{\theta }_0$ denotes the dual with
        respect to the induced metric on $\cal{S}$.
\item[3)] $\omega=\alpha=0 $: The representative in this orbit becomes
        $X=(0,0,0;\theta_0,0)$, which implies
        $\theta_0\d f=0 \Rightarrow f(x)=h(\tilde{\theta}_0x)$ ,
        where the same notation is used as in the cases above.
\item[4)] $\omega\neq 0, \alpha \neq 0$ : This case requires some discussion.
        Using polar-coordinates $(\rho,\phi)$ and the form of the
        representative $X=(1,0,\omega;0,0)$, we find
        $f = \gamma \d _\phi f\quad (\gamma = \frac{1}{2}\epsilon^{ij}
        \omega_{ij}$) $\Rightarrow f(x)= h(\rho) e^{\gamma\phi}$,
        which is not globally defined on $\cal{S}$. Unwrapping $\cal{S}$
        with the help of
        $$
        w=\log z\quad p + iq = \log\rho + i\phi
        $$
        $f(x)$ becomes a well-defined function on the infinite
        cover of $\cal{S}$, i.e. $f(p,q)=h(e^p) e^{\gamma q}$.
\end{itemize}
All of the above cases  possess a four-parametric Killing-algebra.
Imposing the vacuum-equations turns 1) into the AS-geometry, the
arbitrary function $h$ in 2) and 4) become harmonic and
3) reduces to a linear profile thus being equivalent to
Minkowski-space.  Higher symmetry-classes may be obtained by combining
the above classes. However, like in the general case, one
cannot assume that both Killing-vectors (and the corresponding
profiles) are simultaneously in their canonical form. In general one
has to apply an arbitrary rnfp-transformation to the canonical form
of the profile relative to the first Killing-vector, and impose the
second Killing afterwards. Taking the respective stability-(sub)groups
of both Killing-vectors into account it is possible to
simplify the arbitrary rnfp-transformation considerably.
Applying this procedure to 1)--4) it turns out that only the
combination of 2) and 3) yields a non-trivial result.
\begin{align}\label{comb}
&f(x)=h_0 \exp\left(\frac{1}{\theta_0^2} \left(\frac{\theta_0\eta_0}
        {\tilde{\theta_0}\eta_0}\tilde{\theta_0}x - \theta_0 x\right)\right)
        \nonumber\\
&X=(1,0,0;\theta_0,0)\qquad Y=(0,0,0;\eta_0,0)
\end{align}
\noindent
Note however that the symmetries of (\ref{comb}) are not compatible with
Einstein's vacuum equations.
Let us finally summarize the results in the following table:
\newline

\noindent
\begin{tabular}{|l|l|l|l|} \hline
profile $f(x,y)$ &Killing-vectors $\xi$ &r &type \\ \hline\hline
 general & $\xi_1=\d_v$, $\xi_2=2 x\d_v + u\d_x, $
        &3 &abelian \\
&$\xi_3=2y\d_v + u\d_y$&&\\ \hline
 $g(x^2+y^2)$ & $\xi_1,\xi_2,\xi_3,$ $\xi_4=x\d_y-y\d_x$ &4&$E_2\times
        \Bbb{R}$\\ \hline
 $h(y) e^{-x/\epsilon}$
        &$\xi_1,\xi_2,\xi_3,$ $\xi_4 = u\d_u -v\d_v + \epsilon \d_x$ &4
        &$[\xi_4,\xi_1] = \xi_1$, \\
&&& $[\xi_4,\xi_2] = \xi_2 + 2\epsilon \xi_1,$ \\
&&&$[\xi_4,\xi_3] = \xi_3$ \\ \hline
 $h(y)$ &$\xi_1,\xi_2,\xi_3,$ $\xi_4=\d_x$ &4
        &$[\xi_4,\xi_1] = 0$,\\
&&&$[\xi_4,\xi_2] = 2\xi_1$,\\
&&&$[\xi_4,\xi_3] = 0 $\\ \hline
$h(e^{p}) e^{\gamma q}$ &$\xi_1,\xi_2,\xi_3,$  &4 &$[\xi_4,\xi_1] = \xi_1$, \\
&$\xi_4 = u\d_u -v\d_v + \gamma
        (x\d_y-y\d_x)$&&$[\xi_4,\xi_2] = \xi_2-\gamma\xi_3$,\\
&&&$[\xi_4,\xi_3] = \xi_3 + \gamma\xi_2$ \\ \hline
$h_0 e^{\frac{1}{\epsilon_1}(\frac{y}{\epsilon_2}-x)}$&$\xi_1,\xi_2,\xi_3,$
        $\xi_4 = u\d_u -v\d_v + \epsilon_1 \d_x$& 5 &$[\xi_4,\xi_1] =
        \xi_1$, \\
&$\xi_5 = \d_x+\epsilon_2 \d_y $&&$[\xi_4,\xi_2] = \xi_2 + \epsilon_1 \xi_1,$\\
&&&$[\xi_4,\xi_3] = \xi_3, $ \\
&&&$[\xi_5,\xi_1] = 0, $ \\
&&&$[\xi_5,\xi_2] = 2\xi_1, $ \\
&&&$[\xi_5,\xi_3] = 2\epsilon_2\xi_1, $ \\
&&&$[\xi_5,\xi_4] = 0, $ \\ \hline
\end{tabular}

\newpage
\section*{Conclusion}
In this paper we have generalized the classification of pp-waves in terms
of Killing symmetries by allowing the profile function to be a delta-like
pulse. Moreover, no field equations were imposed. The method presented
relies on the analysis of the adjoint orbits of the so called
normal-form-preserving diffeomorphisms. Our calculations show that new symmetry
classes occur even in the vacuum case.
 In a forthcoming paper we intend to give a more detailed presentation for
general wave profiles. There the analysis is considerably more involved.
 Nevertheless, we feel that the method presented may  also be useful for
classifying other types of space-times.

\vfill
\end{document}